\documentclass[preprint,authoryear,12pt]{elsarticle}

\usepackage{amssymb}
\usepackage{graphicx,bm}
\usepackage{float, Abbrev}
\usepackage[fleqn]{amsmath}  
\usepackage{color, url}

\newcommand{\be}{\begin{equation}}
\newcommand{\ee}{\end{equation}}
\newcommand{\ba}{\begin{eqnarray}}
\newcommand{\bmat}{\begin{pmatrix}}
\newcommand{\emat}{\end{pmatrix}}

\newcommand{\arcsec}{''}
\newcommand{\arcmin}{'}

\def\simlt{\lower.5ex\hbox{$\; \buildrel < \over \sim \;$}}
\def\simgt{\lower.5ex\hbox{$\; \buildrel > \over \sim \;$}}

\usepackage{graphicx} 

\journal{Astronomy and Computing}

\begin{document}

\begin{frontmatter}



\title{Observing Dark Worlds:\\ A crowdsourcing experiment for dark matter mapping}


\author[roe]{David Harvey }
\ead{ drh@roe.ac.uk}
\author[ucl]{Thomas D.\ Kitching}
\ead{ t.kitching@ucl.ac.uk}
\author[kaggle]{Joyce Noah-Vanhoucke}
\author[kaggle]{Ben Hamner}
\author[tim]{Tim Salimans}

\address[roe]{SUPA, University of Edinburgh, Royal Observatory, Blackford Hill, Edinburgh EH9 3HJ, UK }
\address[ucl]{Mullard Space Science Laboratory, University College London, Holmbury St Mary, Dorking, Surrey RH5 6NT, UK}
\address[kaggle]{Kaggle, Inc., 188 King St. Suite 502, San Francisco CA 94107 USA}
\address[tim]{Erasmus University, Rotterdam, The Netherlands}

\begin{abstract}
We present the results and conclusions from the citizen science competition `Observing Dark Worlds', where we asked participants to 
calculate the positions of dark matter halos from 120 catalogues of simulated weak lensing galaxy data, using computational methods. 
In partnership with Kaggle (\url{http://www.kaggle.com}), 357 users participated in the competition which saw 2278 downloads 
of the data and 3358 submissions. 
We found that the best algorithms improved on the benchmark code, {\tt LENSTOOL} by $>30\%$ and could measure the positions of 
$>3\times10^{14}M_\odot$ halos to $<5\arcsec$ and $<10^{14}M_\odot$ to within $1\arcmin$. 
In this paper, we present a brief overview of the winning algorithms with links to available code.
We also discuss the implications of the experiment for future citizen science competitions. 
\end{abstract}


\end{frontmatter}

\newpage
\section{Introduction} \label{sec:intro}

Dark matter dominates the mass content of the Universe (see for example Massey, Kitching, Richards, 2012; Amendola et al. 2012 for reviews), in particular on 
galaxy and galaxy cluster scales where the ratio of total mass to observed baryonic matter is a factor of at least $10-100$. 
In fact approximately $30\%$ of the total energy budget of the Universe is in the form of non-baryonic matter \citep{planckpars}.  
Further constraining the nature of dark matter has become one of the most important problems in physics \citep{PeterRev}. 
However, despite the macroscopic total abundance of this non-baryonic component of the Universe being well determined, the understanding of the sub-atomic physics of dark natter is 
not; if indeed dark matter is a subatomic particle at all.  
Under the assumption that dark matter is a non-relativistic particle when it decouples from baryonic, ordinary, matter in the early universe, and that it is collisionless, 
one can qualitatively reconstruct the observed large scale structure in N-body simulations \citep{EvolutionLSS}, with the baryonic physics \citep{BaryonSim} limiting our knowledge at the sub megaparsec scales.
Dark matter is hypothesised to exist in clouds of particles that self-gravitate into bound systems, these clouds are refered to as dark matter `halos' since observationally 
dark matter appears to have concentrations that are highest in clouds that surround baryonic matter.

It is hypothesised, from N-body simulations, that there are several problems with the collisionless dark matter paradigm at small scales, where predictions 
begin to depart from measurements in data. These are the `too big to fail problem' \citep{toobigtofail}, and `the cuspy halo problem' \citep{corecusp}. 
The former refers to the observation that N-body simulations predict far more large sub-halos in galaxies that exhibit star formation than we see in the Milky Way, 
the latter refers to the observation that galactic halos have `cores' (a high density of dark matter) which are inconsistent with those predicted by N-body simulations \citep{hierarchial}.
In order to reconcile these inconsistencies, one can invoke a variety of mechanisms that add complexity to the collisionless dark matter scenario (so called Cold dark matter or CDM paradigm). 
For example warm dark matter, self interacting dark matter (SIDM), and the impact of baryons on CDM all have the potential to account for the observed differences 
\citep{ObserveSIDM, GalaxySIDM}, or N-body simulations are not representative of the Universe in some other respects. 

From observations it has been observed that the highest ratio of mass-to-light, i.e. the largest concentrations of dark matter, are in galaxy clusters. 
These clusters are therefore the best available astronomical `laboratories' to study the properties of dark matter because not only is there a relative overabundance, 
but there is also a relatively large amount of baryonic matter against which dark matter properties can be calibrated and compared. Previous work studying the 
distribution of dark matter in galaxy clusters has led to discoveries of colliding clusters and evidence of dark matter 
\citep{bulletclusterA,separation,bulletclusterB,minibullet,A2744,musket,A520,A520A,A520B}. 

In this paper we focus on the technique of gravitational lensing as a probe of the dark matter distribution. 
According to general relativity, the presence of mass acts to distort the path of photons through the Universe relative to the path that would have been take in the 
absence of mass \citep{BS01,RefregierRev,HoekstraRev,MKRev}. 
Gravitational lensing therefore, probes the total mass along the path of a photon and, because our Universe is dominated by dark matter, has become the primary technique for mapping 
dark matter. The ability to independently measure the distribution of the total matter content, without some assumed relation between observed galaxies and the underlying 
gravitational potential means that gravitational lensing is less sensitive to potential astrophysical systematics. 
In galaxy clusters, in the regime where the lensing mass is large, gravitational lensing effects can result in multiple images of galaxies and highly distorted images; so called 
strong lensing. However every galaxy is lensed by some amount; an effect 
that does not result in multiple images or strong distortions 
but only causes a change in the observed ellipticity of the source galaxy: so called `weak lensing'. The small change in ellipticity caused by weak lensing is refered to as `shear'. 

In this paper we will present the analysis of simulated weak lensing data around simulated galaxy clusters. The analysis of these simulations, in an effort to improve the algorithms 
that are used to infer the mass distribution from weak lensing data, were used to define a citizen science competition that was crowdsourced to the public. 

\subsection{Standard approaches to dark matter reconstruction}

The fidelity with which algorithms are required to map the dark matter distribution in galaxy clusters depends on the range of scales in question. 
Although it is possible to map the distribution of matter using galaxy velocities it has become increasingly popular to use gravitational lensing 
to determine the total matter distribution. There are several approaches that have been developed within the field of weak lensing where algorithms 
are split mainly into two categories based on the type of model used:
\begin{itemize}
\item Parameteric methods involve fitting a physical model to the data and constraining a number of 
parameters in that model. 
\item Non-parametric methods attempt to directly convert from the measured shear to some projected mass density. 
\end{itemize}
For a recent review of the standard approaches see \citet{ReconRev}. 

Throughout this paper we shall refer to the benchmark code {\tt LENSTOOL}.
{\tt LENSTOOL} \citep{lenstool} is a public strong and weak lensing gravitational mass reconstruction method that fits dark matter halos, parameterised by a parametric radial profile,
 to data and determines posterior probabilities for the parameters via a Bayesian sampling method. 
Given $50\arcsec$ priors (not applied to this competition), \citet{Harvey13} found that the accuracy of {\tt LENSTOOL} is roughly $\sim10\arcsec$ for a halo of mass $\sim10^{13}M_\odot$, and is robust to most potential systematics involved in parametric fitting. This code was run on the competition and presented in this paper in order to provide benchmark analysis on individual scores.

\subsection {Expert citizen science}
Citizen science has recently become a productive tool in the analysis of large complicated databases for which algorithms are unable to provide reliable results. 
Pioneering this work in science is the Zooinverse\footnote{\url{https://www.zooniverse.org}}. The Zooniverse is a database of various projects including (amongst others), Moon craters, whale sounds and galaxies.
In each case, a sample of images/sounds or other data is presented to a user (a `citizen'), 
who is then guided through steps to classify that sample into a particular category based on their personal 
judgement. In many cases, such as the identification of complex galaxy morphologies, human-based classification is more reliable than current automated algorithms. The 
science is achieved through the statistical analysis of the human-classified data sets.  

The success of using humans to classify large databases of complicated objects relies on the number of humans doing the classification to be large, to avoid 
individual subjectivity (although there are common inter-subjective biases in human object recognition that need to be found and quantified). 
The advantages of using a large population, a `crowd', to solve or `source' a classification problem is currently refered to a `crowdsourcing'. 
However there are two regimes in which the human-classification mode of crowdsourcing a problem is limited
\begin{itemize}
\item
When the data set, or the number of classification categories, becomes too large for a population of humans to analyses in a reasonable time period. An example would be a database of 
several billion astronomical objects, each of which needed many minutes of classification.
\item 
When the precision required for a measurement is very high. An example would be in weak lensing measurements where the accuracy required is sub-percent in the bias 
of ellipticity measurements of galaxies. 
\end{itemize}
In these regimes algorithms are required to analyse the data. However the crowd can still be used, but in a different mode: instead of classifying, the crowd can be asked to write 
computer algorithms to solve the task at hand. In this regime one needs to set the problem to a targetted group of computer programming literate individuals or teams, with sufficient 
motivation (either in the form of a prize for writing the best algorithm or other), and with a clearly defined objective measure for what is meant by the best algorithm, i.e. a metric 
for success. The algorithm-writing mode of crowdsourcing in astronomy has only recently been utilised for example in a competition \citep{MDM}, in partnership 
with Kaggle\footnote{\url{http://www.kaggle.com}}, where the problem of weak lensing shape measurement was set to the public (see also 
Heymans et al. 2006; Massey et al. 2007; Bridle et al., 2010; Kitching et al, 2012; Mandelbaum et al. 2013 for more complex challenges in the same area). 

In this paper we present the results of 
crowdsourcing the problem of using weak lensing measurement to create maps of the dark matter distribution around galaxy clusters. `Observing Dark Worlds' was a competition in partnership with Kaggle, whereby we asked participants to reconstruct the positions of simulated dark matter halos in fields of galaxies. By varying the parameters of the fields, such as the mass and the galaxy density, we aimed to probe the sensitivity and behaviour of the reconstruction algorithms. In a bid to develop an algorithm that was systematically unbiased and statistically precise, we supplied 120 clusters ranging in mass from the group scale ($10^{13}M_\odot$) to super-cluster scale ($10^{15}M_\odot$).

This paper is organised as follows, in Section \ref{sec:overview} we will outline the premise of the competition, including a description of the data provided to participants. In Section \ref{sec:results} we present our results. Section \ref{sec:algorithms} gives a detailed description of the winning three algorithms and in Section \ref{sec:conc} make our conclusions.

\section{Observing Dark Worlds}\label{sec:overview}

Determining the distribution of dark matter  in galaxy clusters has generally been focused on how well one can reproduce the macroscopic properties of dark matter including the mass and concentration 
parameter \citep{strongweak, strongweakA,strongweakB,strongweakC} . \citet{bulleticity} and later \citet{Harvey14}, developed a method to constrain the self-interaction cross-section of dark matter using the position of dark matter substructure. In order avoid systematic errors, this technique required accurate estimation of substructure and systematic errors to be $<0.5\arcsec$ \citep{Harvey13}. Any biased estimate of position of galaxies would result in a spurious constraint of SIDM.
With this in mind our aim for the competition `Observing Dark Worlds' (ODW) was to encourage the development of new algorithms to reconstruct the 
position of dark matter halos in galaxy clusters with a systematic bias $<0.5\arcsec$.

\subsection{The Competition}
In order to achieve our competition aim, we required competitors to reconstruct the positions of dark matter halos in a number of simulated galaxy clusters with varying parameters. We provided users 
with data to `train' on and a test set on which they submitted blind answers. Participants had two months to improve their algorithms and make submissions to the Kaggle website. 
After two months, on the proviso that they provided the correct documentation concerning their algorithm, the top three participants received a reward: in the case of ODW this was generously provided by Winton 
Capital Management\footnote{\url{https://www.wintoncapital.com}}.

\subsection{The Data}

The competition consisted of three different types of data sets (this is similar for the majority of Kaggle competitions);
\begin{description}
\item[\textbf{Training Set:}] The set on which users could train their algorithms. Users had access to the galaxy catalogues that included positions and ellipticities in the 
field \textbf{and} the true positions of each dark matter halo.
\item[\textbf{Public Test Set:}] The set on which users were tested. They had access to catalogues of galaxies which contained positions and ellipticities, however the positions of the dark matter halos were unknown. Users had to submit their predictions to the Kaggle website and were scored according to some metric (see section \ref{sec:metric}) and the result published on a leaderboard. 
However the final results \textit{were not} based on these results.
\item[\textbf{Private Test Set:}] Another, separate set of data on which users were be tested. Similar to the public test set, users were required to submit predictions for the positions of the dark matter halos in these fields, however they did \textit{not} receive any feedback on their score via the live leaderboard. These scores were kept secret (private) until the competition finished at which point they were 
revealed and the final result was be based on these. This is designed to prevent people over-fitting to the data. Note that competitors did not know which galaxy clusters were in the private or public test set.
\end{description}
Typical training sets supplied for machine learning problems usually incorporate a large number of training samples (many more than the test sets), from which a computer program can `learn', 
and then a test set for which competitors have to submit their predictions to a blind sample. However such training sets may not be possible in real world astronomical situations. 
In the decision for the set size for the training and test we considered;
\begin{itemize}
\item What sized data set well reflects what one would see in real observational data?
\item Do you want to place emphasis on quick accurate algorithms, or complex precise algorithms?
\end{itemize}
The latter was an important factor when we considered the sample size since we did not want to limit the complexity/run-time of algorithms. Currently {\tt LENSTOOL} takes $\sim30$minutes/cluster, 
which is acceptable for samples of $\sim100$ clusters (on a current machine). 
The number of clusters that one asked users to use was inevitably going to affect the type of algorithm submitted throughout the competition. 
Moreover, the properties of the clusters themselves affected the set sizes since if one requested many halos per cluster then one was requiring users to constrain many more parameters. 
In addition, small mass halos (or a high un-sheared/intrinsic ellipticity noise), which are harder to constrain, may require many likelihood evaluations during fitting. 
We therefore considered; the number of dark matter halos in each cluster, 
the properties of the halos in each cluster including; mass, profile, shape and concentration and their distribution in the field, the number of background lensed galaxies per cluster and the noise on the galaxies. Taking this also into account we choose the following set of parameters: 
\begin{itemize}
\item 360 training halos and 120 test clusters, of which three quarters are the private test set and one quarter is the public test set. 
\item For each set exactly one third of the clusters have one, two and three halos in them and are randomly distributed in the field.
\item Each cluster has one main halo which has a mass randomly chosen between $1-10\times10^{14}M_\odot$. Where there is greater than one halo in the cluster the second (and third) halo mass is randomly selected between  $1-10\times10^{13}M_\odot$ imitating an in-falling galaxy group.
\item To reflect the field of view and depth of the Hubble Space Telescope and the typical intrinsic ellipticitiy from the COSMOS field \citep{COSMOSintdisp}, we randomly select the source galaxy density between $30-80$ galaxies/arcmin$^2$ and apply a intrinsic ellipitciity with a mean, $\langle \epsilon_{\rm int} \rangle=0$ and a dispersion, $\sigma_\epsilon=0.3$ and place them randomly in a field of $3\times3$ arcminutes.
\item Finally, we told the competitors exactly how many halos were in each field.
\end{itemize}
This is shown in more detail in Table 1.
\begin{table}\label{tab:pars}
\begin{centering}
\begin{tabular}{| c | c |}
\hline
\multicolumn{2}{|c|}{Data Set Parameters} \\
 \hline
\# Clusters [Training, Public, Private] & [360, 30, 90]  \\
1 Halo [Training, Public, Private]& [120, 10, 30] \\
2 Halo [Training, Public, Private] &  [120, 10, 30]  \\
3 Halo [Training, Public, Private]&  [120, 10, 30]   \\
NFW [Training, Public, Private] & [180, 15, 45] \\
SIS [Training, Public, Private] & [180, 15, 45] \\
1st Halo Ellipiticity & [0, 0.3] \\
1st Halo Mass & $[1,10]\times10^{14}M_\odot$ \\
2nd Halo Mass & $[1,10]\times10^{13}M_\odot$ \\
3rd Halo Mass & $[1,10]\times10^{13}M_\odot$  \\
Intrinsic ellipticitiy dispersion& 0.3 \\
Galaxy Density & [30,80] galaxies/arcmin$^2$ \\
Field of View & $3\times3$ arc minutes \\
\hline 
\end{tabular}
\caption{The parameters used for the data sets provided for the ``Observing Dark Worlds'' competition.}
\end{centering}
\end{table}

\subsection{ The Metric}\label{sec:metric}
One of the most important aspects of a competition such as ODW is how one measures the success of an algorithm. 
A good metric should accurately reflect how well an algorithm is doing at achieving the aims which you have set. 
In other words, to what extent has an algorithm achieved your required goals? The consequences of this are that the design of metric needs to be done meticulously, such that it rewards and penalises aspects 
of the problem that you are interested, and not interested in, respectively. Specifically in the context of this competition, the metric was required to achieve the aims set out 
in Section \ref{sec:overview}. To this extent, we required a metric that not only rewarded competitors for providing solutions that were close to the halo positions, but also rewarded solutions that were 
not systematically biased in a particular angular direction, i.e. an unbiased solution would be angularly invariant on average to the true position of the halo. 
We therefore constructed a metric of two parts, a distance part, $F$, and an angular part, $G$ and combined them such that the overall score, $m$, was,
\be
m = \frac{F}{1000}+G,
\label{eqn:metric}
\ee
where we weighted it such that {\tt LENSTOOL} achieved a score of approximately $1.0$.

\subsubsection{The Distance, F Metric}
The first part of the metric quantified the ability of an algorithm to produce halo positions that were as close as possible to the true positions. 
However, one issue we met was that in the case of clusters with more than one halo, it was not always clear which halo users were predicting, i.e. if they had named their first prediction, ``halo1'', 
it was ambiguous to which halo this referred to. We therefore selected the halo pairs (users halos to true halos) 
such that it optimised the distance part of the metric. The `F' part of the metric was in the form,
\be
F= \sum_{k=0}^{\rm nClusters}{\rm arg~min}\left\{ \sum_{i=0}^{\rm nHalos} \sqrt{(x_{ik}-x_{jk})^2+(y_{ik}-y_{jk})^2}\right\}_j^{\rm config}\
\label{eqn:fmetric}
\ee
where the min function is over all the pair configurations in the cluster, where for two halos there was two and for three there was six. We normalised the $F$ metric to provide an 
approximately unit score for {\tt LENSTOOL}.
\begin{figure}[h]
		\begin{centering}
	 		\includegraphics[width = 8cm]{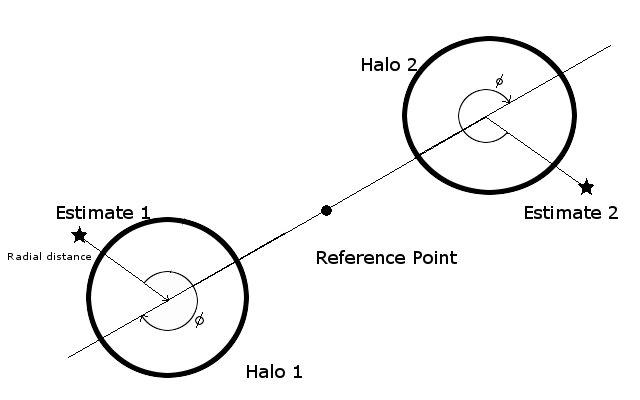}
			\caption{\label{fig:fmetric}
The angular, G metric. $\phi$ was calculated as the angle between the the vector jointing the true and users predicted position and the vector joining the centre of mass of the system (reference point) and the true position.}
			\end{centering}
\end{figure}

\subsubsection{The Angular, G Metric}
In any competition, all participants will attempt to win, which will mean that they will tune their algorithms such that they achieve the best metric score possible. In the case 
where $m\propto F$, participants could have continued developing their algorithm such that they reduced the separation between predicted and true position of halos. 
Although this may have resulted in predictions that are very close to the true halo positions, it would not have solved the second aim of developing an algorithm that is systematically unbiased. 
We therefore included a second part to the metric that rewarded angular invariance of predictions. 

To calculate the angular invariance, we derived the angle of the vector between the true and the predicted position with respects to the vector that connected the true position and 
the centre of mass of the system (or reference point as it was denoted). Once we had this angle, $\phi$, we took the average vector that this angle defined in a unit circle, i.e.
\be
G=\sqrt{\left(\frac{1}{N}\sum_{i=0}^N \cos(\phi_i)\right)^2+\left(\frac{1}{N}\sum_{i=0}^N \sin(\phi_i)\right)^2},
\ee
where the sum $N$ is over all halos in all clusters. Figure \ref{fig:fmetric} shows diagrammatically how we calculated $\phi$.

\section{Results}\label{sec:results}

\begin{figure}
		\begin{centering}
	 		\includegraphics[width = 8cm]{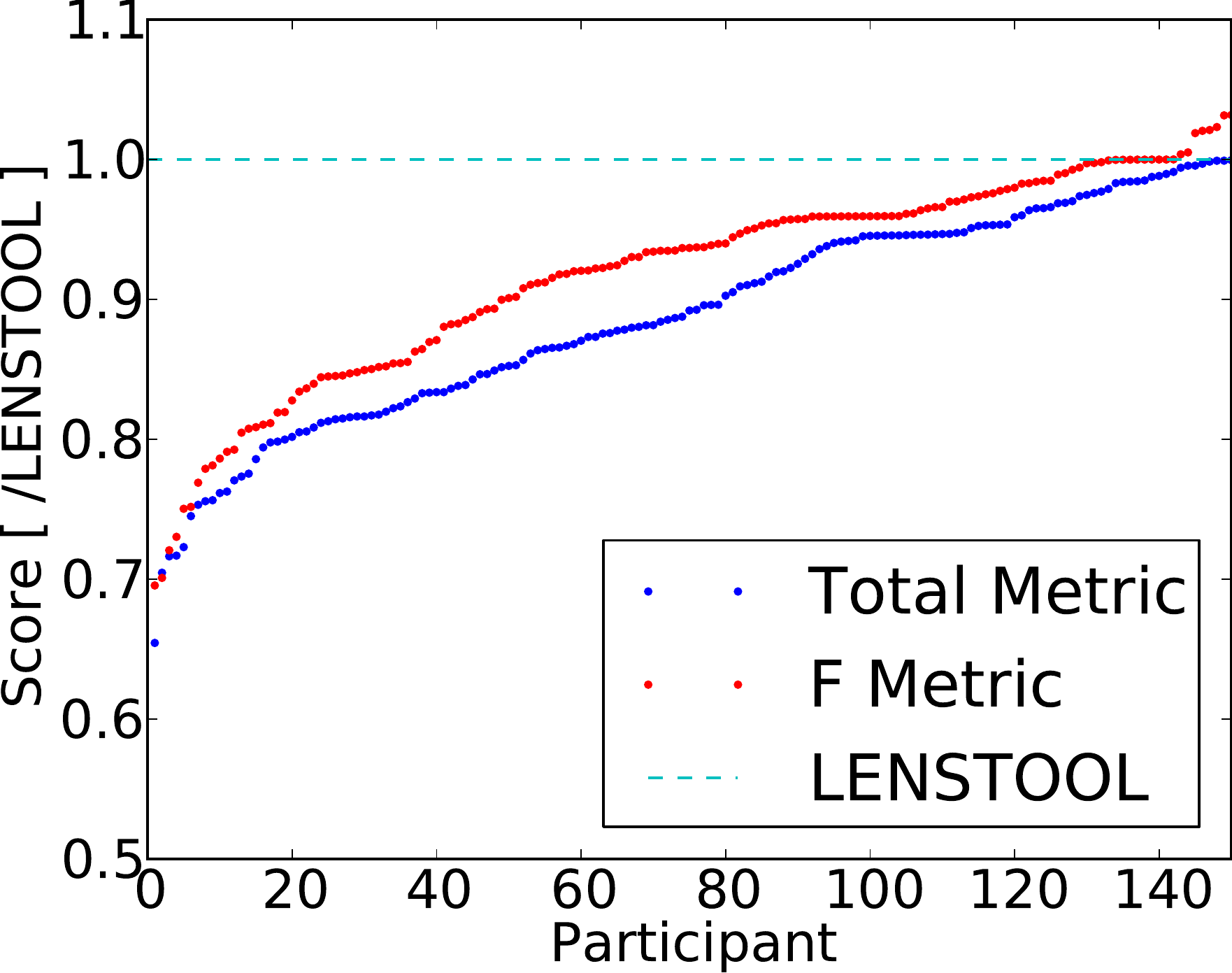}
			\caption{\label{fig:results}
The figure shows the score of the top 150 competitors normalised to the score of {\tt LENSTOOL} for easy direct comparison (where {\tt LENSTOOL} $=1.0$) as a function of the participants final 
leaderboard rank. The blue points refer to the total score of the competitor and the red the radial part. The scores have been sorted and therefore the points in the radial do not directly correspond to the points in the blue.}
			\end{centering}
\end{figure}

Here we present the results from the ODW competition. 
We present the original metric, and also results that were calculated using only the distance part of the metric, which we found to dominate the total metric. 
Figure \ref{fig:results} shows the results for the two metrics normalised to the score of {\tt LENSTOOL}, (in other words a score of 1.0 is the same score as {\tt LENSTOOL}), as a function of 
competitor for the top 150 competitors. We have sorted the results such that the total score does not correspond with the radial score of the same participant. 

From the initial results we find that of the 357 participants, 143 had a score better than the {\tt LENSTOOL} benchmark with the top 27 competitors registering better than a $20$\% improvement, and the top 
competitors recording a $>30\%$ improvement. We find 
that the act of removing the angular part of the metric slightly reduces the relative score of the competitors with respects to {\tt LENSTOOL}, with 150 of the 357 competitors (42\%) achieving a 
better score than {\tt LENSTOOL}, and that the top competitors still recorded a $>30\%$ improvement.

\begin{figure}
		\begin{centering}
	 		\includegraphics[width = 16cm]{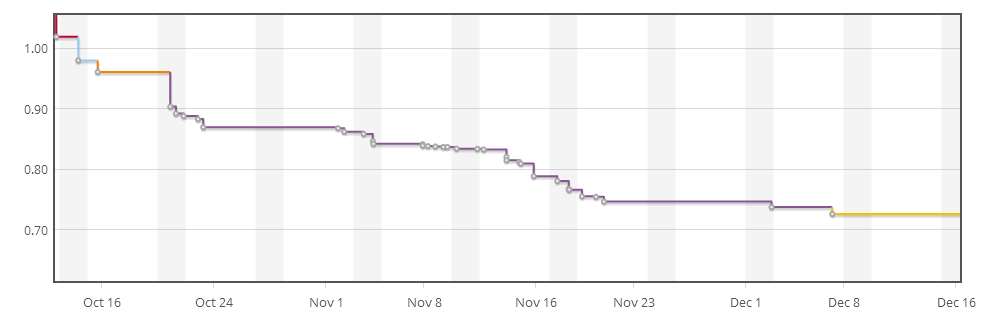}
			\caption{\label{fig:timeline}
Timeline for the improvement of the best scoring algorithm. Initially, a lot of progress is made, however towards the end, progress plateaus with few and small incremental improvements. Although typical, this competition saw more more regular improvements than typical machine learning problems.}
			\end{centering}
\end{figure}

Figure \ref{fig:timeline} shows the incremental improvement of the best scoring algorithm with time. It can be seen that in the initial periods of the competition regular, large improvements are made, and as the compeition continues these improvements become less frequent. This is typical of a machine learning competition, however ODW saw slightly more regular improvement and increased late time large improvements than a typical competition.

\begin{figure}
		\begin{centering}
	 		\includegraphics[width = 14cm]{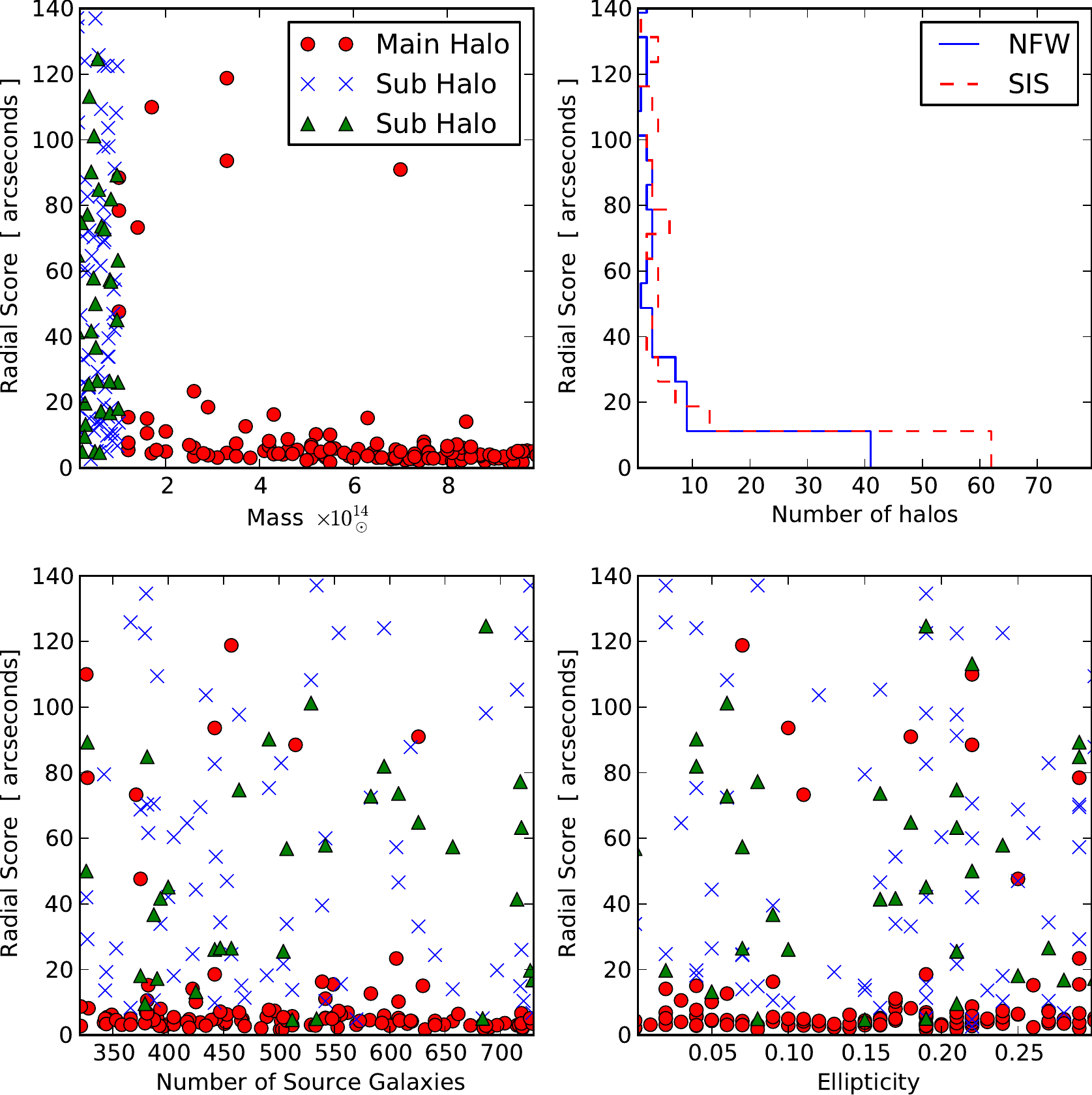}
			\caption{\label{fig:varComp}
The figure shows the average score of the top 150 competitors radial score as a function of various competition parameters. It is clear that there is an obvious, dominant trend with respect to the mass of the halo. Also the top right panel shows how more SIS halos were constrained to within $15\arcsec$ than NFW.}
			\end{centering}
\end{figure}

Figure \ref{fig:varComp} shows how the average of the top 150 competitors performed as a function of various components of the competition. We find that the mass of the halo is the 
dominant variable in the estimation of the position of dark matter halos, and the methods were unaffected by the number of source galaxies. Interestingly competitors did better when the halo was a 
Single Isothermal Sphere and not an NFW profile. We hypothesise that this is because an SIS is peakier, with no core, and therefore estimating its peak is easier. 
We find a weak trend between the ellipticity of the main halo and the position estimates. 

\begin{figure}
		\begin{centering}
	 		\includegraphics[width = 14cm]{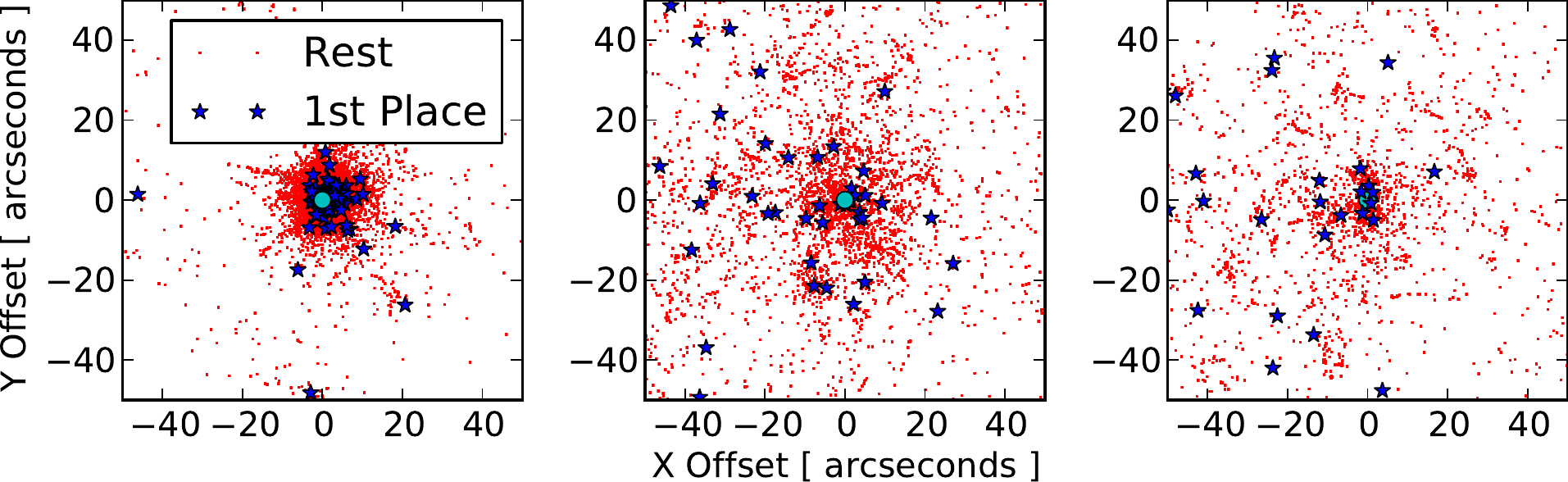}
			\caption{\label{fig:PicPos}
The best radial position of each halo for the top 150 competitors. The first panel shows the results of the main halo in each case, and the second two panels, the sub halo. The blue dots in each case represent the winners estimated positions. One can clearly see that the points are much more clustered for the main halo.}
			\end{centering}
\end{figure}

Figure \ref{fig:PicPos} shows graphically the distribution of estimates of the 150 participants. We can clearly see that in the left hand panel, that shows the main halo estimates, positions are 
much more concentrated and clustered about the true position than the sub halos shown in the second two panels. First place submissions are shown as blue stars and the cyan dot is the truth.

\begin{figure}
		\begin{centering}
	 		\includegraphics[width = 8cm]{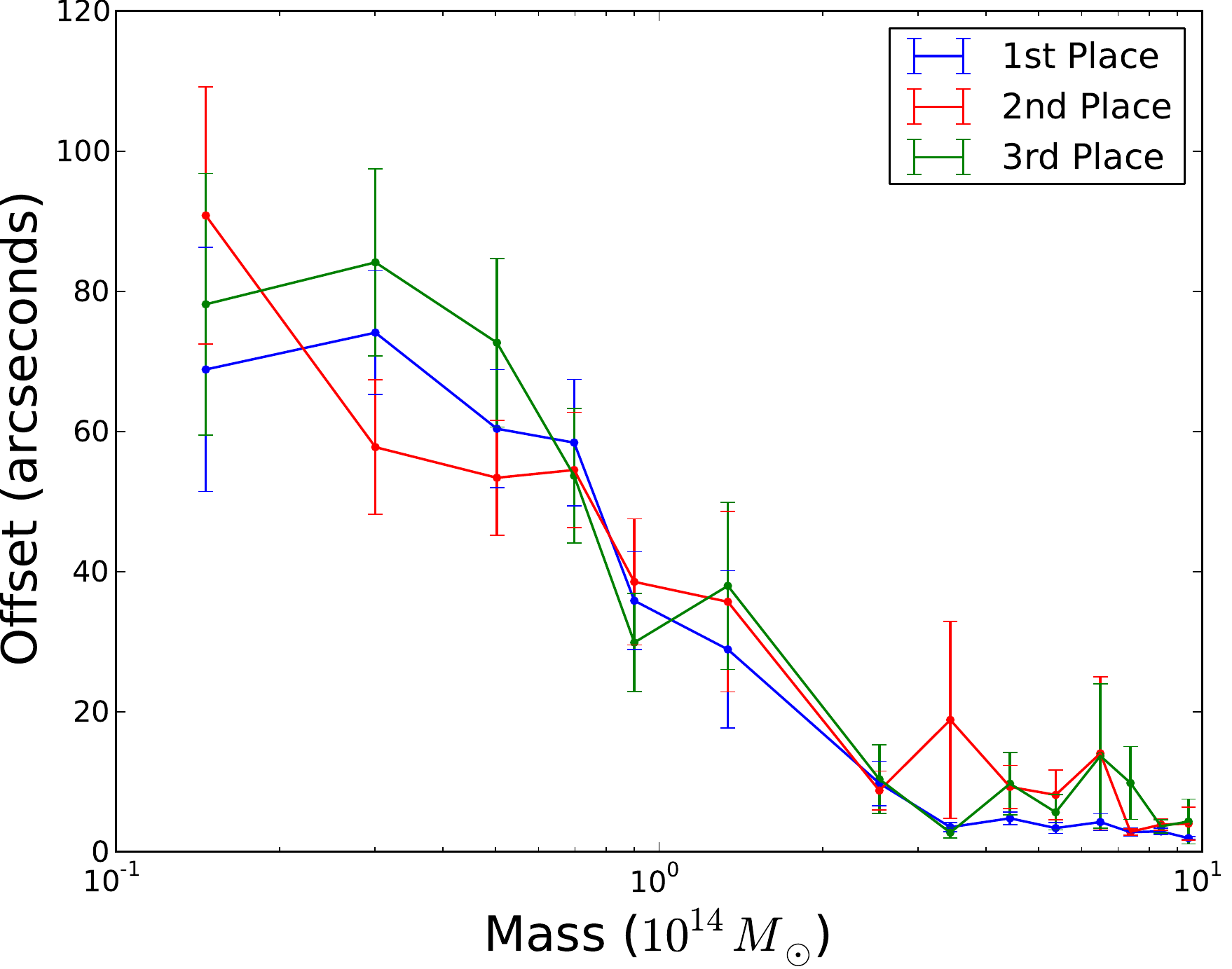}
			\caption{\label{fig:topSub}
The radial distance the top three competitors are from the true position in arcseconds. The submissions have been binned by mass, and the error bars are the error in the mean of the radial distances in the mass bin.}
			\end{centering}
\end{figure}

In order to observe the statistical accuracy of the best submissions, we binned up the submissions in mass. Figure \ref{fig:topSub} shows the results of the top three placing algorithms. The error bars show the error in the mean radial position for that particular mass bin. We find that the best algorithms could constrain the large mass halos to $<5\arcsec$ rising to $\sim60\arcsec$ for smaller mass halos. 

\subsection{Metric Stability}
In order for the competition to provide accurate feedback to participants, the metric quantifying this quality needs to distinguish between two algorithms that are very similar,
and determine which is better at achieving your research goal. There are many ways that one can reduce the noise in the metric and ensure the correct algorithms win.

\begin{figure}
                \begin{centering}
                        \includegraphics[width = 9cm]{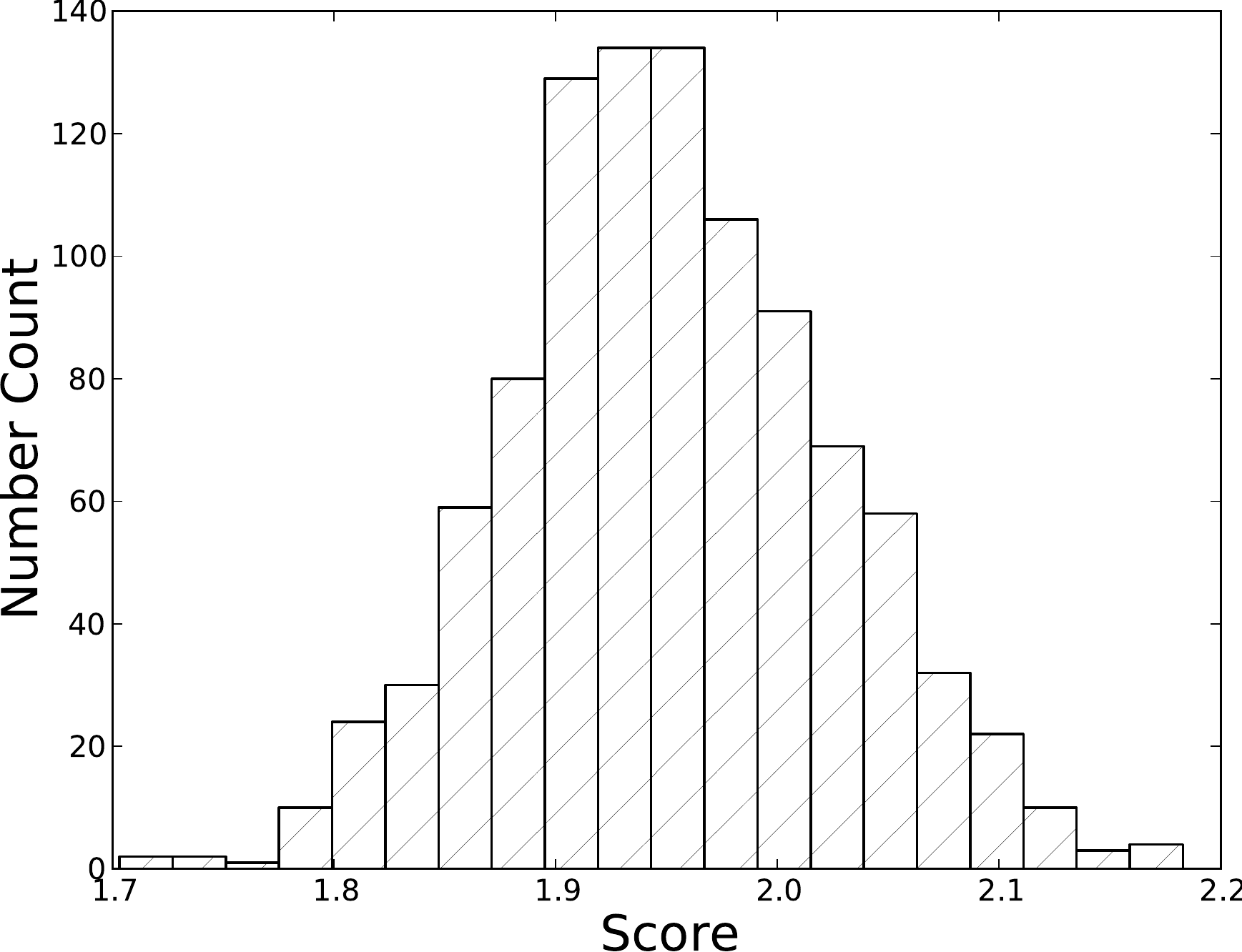}
                        \caption{\label{fig:metricStab}
The score of a random sample of fake entries; 1000 fake submissions with random guesses were scored.  This noise on the metric
lead to less efficient feedback from the leaderboard given to participants.}
                        \end{centering}
\end{figure}

The first method to minimise the noise is to have a well defined metric that is simple and has minimal intrinsic variability.
The metric set out in this competition had two parts; the distance part and the angular part. The former was a direct probe for
the precision of an algorithm, and the second part was used to determined any systematics in the algorithm. However the angular part of the metric was also intrinsically noiser than the radial part.
Figure \ref{fig:metricStab} shows the variance in a random set of submissions, whereby each halo was a random guess within the field of view of the cluster.
It can be seen that there was a large variance in scores due to the nature of the metric.
In the limit that the number of clusters tends to a large number, noise will average out and the score on the metric will tend to truly reflect quality of the algorithm.
However in a bid to reflect the expected number of clusters from typical data sets in the near future we only provided a small number of clusters.
Moreover, we did not want to limit participants to algorithms that had requirements of $>1$ second/cluster reconstruction times.
In normal machine learning competitions, data sets are of the order $10^5$ samples, and in the scenario where a lot of people enter with similar algorithms, the scores at
the top of the leader board can be separated by $\sim 0.01\%$. This requires the noise to be another order of magnitude less than this.

Another consequence of small training and test sets is the effective Poisson noise in the sampling of parameters for the data.
We sampled each parameter for the field in Table 1 from a uniform distribution. As a result the mass of the halos in each test set varied by a large amount,
The public test set consisted of smaller sub halos than the private test set. In this sense, the public test set was not only harder, but below a certain halo mass the noise in the
positions was larger, making the public leaderboard noiser than the private leaderboard; this is something that should be addressed in designing future competitions. For a full description of what considerations should be taken into account when designing a research competition see \citet{competitions}.

\section{Successful Algorithms}\label{sec:algorithms}
The winning solutions all had similar Bayesian approaches. As part of the competition, a requisite for claiming the prize was that both the code and a detailed description of its working was required. 
As such, we now present the main properties of the winning solutions. Note that since these models were presented as non-astronomically motivated algorithms (the motivation was to win the competition), 
many of the parameters used have no physical units and were based on the experience of the data. These are brief summaries of the solutions, and we encourage the reader to use the embedded URL's to find
more information. 

\subsection{First place: Tim Salimans}
This was a Bayesian solution\footnote{\url{http://timsalimans.com/observing-dark-worlds/}} to the problem and was made publicly 
available\footnote{\url{http://timsalimans.com/code-for-the-dark-worlds-competition/}}. There were four main steps in this code;
\begin{enumerate}
\item 
Initially derive some prior information, $p(x)$, on the positions of the dark matters halos prior to looking at the data. This was assumed to be flat across the field of view.

\item 
From the data, formulate some model for the likelihood of a dark matter halo position given some galaxy information, $p(e|x)$,
\be
p(e_i|x) = N(\sum_{j=0}^{\rm all~halos}d_{i,j}m_jf(r_{i,j}),\sigma^2),
\ee
where the probability is assumed to be a normal distribution, $N$. 
$d_{i,j}$ is the tangential direction to the vector joining halo $j$ to galaxy $i$, (i.e. the direction in which the galaxy is lensed), $m_j$ is 
the mass of the the halo, $r_{ij}$ is the euclidean distance between the halo centers and the galaxies and $f(r_{ij})$ is a radial profile of the halo.  In order to determine the model settings he first placed priors on the masses, $p(m|q)$, where $q$ is the set of parameters determining the shape of the prior and the functional form of $f(r;w)$, with $w$ parameterizing $f()$, he then calculated the marginal likelihood of the ellipticities on the training data (since there the positions of the halos are known): 
\be
p(e|q,w,\sigma^2) = \int p(e|f(r;w),m,\sigma^2)p(m|q) dm,
\ee
By maximizing the marginal likelihood with respects to q, w, $\sigma^2$, he got an efficient and consistent estimates of the true profile parameters and priors, assuming the model is correct. From this he fixed the dispersion of the likelihood model at $\sigma^2=0.05$. The main halo mass was selected from a log uniform distribution between $40$ and $180$ and the sub-halo mass was fixed to be $20$, i.e. $M_\mathrm{S}/M_\mathrm{M}=[0.11,0.5]$. The halo profiles were a simple inverse distance profile with a core,
\be
f(r_{i,j},r_\mathrm{c})=\frac{1}{\max(r_{i,j}, r_\mathrm{c})},
\ee
where the core radius, $r_\mathrm{c}$ was fixed at $r_\mathrm{c}=240$ and $r_\mathrm{c}=70$ for the main and sub halos respectively.

\item 
Using Bayes theorem, calculate the probability of a dark matter halo position, given some galaxy information via 
\be
p(x|e)=\frac{p(e|x)p(x)}{p(e)}
\ee

\item 
Finally minimise the ODW metric with respects to the parameters of the dark matter halos, i.e.
\be
\hat{x} = {\rm arg~min}_{\rm q}\mathbb{E}_{p(x|e)}L({\rm q}, x),
\ee
where q are the predictions of the positions of the dark matter halos. This was implemented via a Monte Carlo Markov Chain (MCMC) Metropolis Hastings sampler, and a simple gradient descent. 
The sampler was also restarted at random points to avoid local minima.
\end{enumerate}

\subsection{Second place: Iain Murray}
In a similar fashion, second place solution also followed a Bayesian solution and the code is also publicly available\footnote{\url{http://blog.kaggle.com/2012/12/19/a-bayesian-approach-to-observing-dark-worlds/}}. Using the same halo profile, this algorithm used slice sampling instead of Metropolis Hastings, as Salimanns used. Since it was assumed that this radial profile was not quite accurate, 
the variance inside the core radius was increased to account for this lack of knowledge. The expected loss with respect to the ODW metric was then minimised. 

\subsection{Third place: Ana Pires}
The third solution was slightly different from the previous two. Pires assumed the halos were sized such that there were (in the case of three halos) a large, medium and small halo. 
This approach also defined a distinct ``elliptical distance'' between the $i$th galaxy, $k$th halo,
\be
d_{i,k} = \sqrt{(x_i-x_k)^2+r_{\sigma,k}(y_i-y_k)^2+2\rho_k(x_i-x_k)(y_i-y_k)}
\ee
where $r_{\sigma,k}$ and $\rho_k$ are free parameters and represent the ratio of the absolute strength of the shear in vertical, $y$ and horizontal, $x$ directions and $\rho_k$ represents a ``distortion''. 
The motivation for this was to represent the distances between halos and galaxies not as ``geometric'' distances, but ``physical'' distance, such that two galaxies may be 
at different geometrical distances from the centre of the halo but experience the same distortion (due to dark matter halo ellipticity for example).

Using the definition for elliptical distance, a model was constructed for the two components of ellipticity, $\chi_1$ and $\chi_2$ for the $ith$ galaxy;
\be
\chi_{1i} = -\sum_{k=0}^{\rm nHalos}\cos(2\phi_{i,k})f_k(d_{i,k})+\alpha_0+\epsilon_{1i}
\ee
and
\be
\chi_{1i} = -\sum_{k=0}^{\rm nHalos}\sin(2\phi_{i,k})f_k(d_{i,k})+\alpha_0+\epsilon_{1i},
\ee
where $\epsilon$ refers to the intrinsic ellipiticity of the galaxy with some variance $\sigma^2$ and mean, $\langle \epsilon\rangle = 0$, $\alpha_0$ is some free parameter and the function $f$ was given by
\be
f_k(x)=\alpha_ke^{-\beta x}
\ee
where $\alpha_k$ and $\beta$ are further free parameters. In order to estimate parameters, she constructed a maximum likelihood criterion, $\eta$, such that
\be
\eta = \mathrm{arg~min}\left(\sum_0^{\rm nSkies}h(\chi_{1i}-\hat{\chi}_{1i};~0.6)+
\sum_0^{\rm nSkies}h(\chi_{1i}-\hat{\chi}_{1i};~0.6) \right)
\ee
where
\be
h(x; \delta)=
\begin{cases}
x^2, & \mbox{if}~|x| < \delta \\
\delta^2, & \mbox{if}~|x| \ge \delta.
\end{cases}
\ee
$\eta$ was then solved via the Nelder-Mead method to derive optimal parameters for the fields.

\section{Conclusions}\label{sec:conc}
We have presented the results of the expert crowdsourcing competition `Observing Dark Worlds'.
The competition was designed to develop weak gravitational lensing algorithms to to reconstruct the position of the peak positions of dark matter halos in galaxy clusters. 
We found that,  of the 357 participants that competed in `Observing Dark Worlds', 150 scored better than the {\tt LENSTOOL} benchmark with the top 27 competitors registering better 
than a $20$\% improvement; and the best algorithms registering a $>30\%$ improvement.  

Notably, the top two algorithms were similar Bayesian fitting methods that fitted various functional forms and then minimised with respects to the Observing Dark Worlds metric. 
We found that the top three algorithms could constrain the large mass halos ($\sim3-10\times10^{14}M_\odot$) to within $<5\arcsec$, which is an accuracy required to used dark matter peak positions to measure
dark matter cross section \citep{Harvey14}. However for smaller mass ($<10^{14}M_\odot$) this rose to $>60\arcsec$. We conclude that these algorithms performed significantly 
better than the public code {\tt LENSTOOL}, however to make a direct comparison further tests will be required on both the winning algorithms and {\tt LENSTOOL} on more realistic simulations, and data.

We found that competitions such as these are a useful approach to developing computer algorithms within the astronomical context. 
Using expert data scientists we found that they can make significant improvements in a relatively short amount of time.

\section*{Acknowledgments} 
We especially thank Kaggle for funding an internship for DH at the company, and for providing advice, guidance and expertise 
on all aspects of this work. We especially thank Winton Capital Management for their generous funding and enthusiasm for physics 
research. The authors are pleased to thank Phil Marshall, Andy Taylor and Richard Massey for many useful conversations and advice.
DH is supported by an STFC studentship. TDK is supported by a Royal Society URF. We would like to also thank Dr. Ana Pires and Dr. Iain Murray for their contributions to the competition.

\section*{References}
\bibliographystyle{model2-names}
\bibliography{bibliography}






\label{lastpage}
\end{document}